\documentstyle[version2,aps]{revtex}
\begin{document}
\begin{title}
Fundamentals of Relativity Providing Time, Distance and Velocity Standards
\end{title}
\author{Ruvin Ferber}
\begin{instit}
Department of Physics, University of Latvia, 19 Rainis Blvd., LV-1586 Riga, Latvia
\end{instit}

\section{Introduction}

Contrary to popular belief, physics is an untidy discipline.
When physicists are faced with the problem of explaining a phenomenon 
which does not
follow from any accepted theory, they raise that phenomenon to the status 
of a postulate, or a principle.
As a result, the foundations of physics present quite a number 
of principles, or postulates,
which are considered as fundamentals for different areas and approaches 
in physics.
One can mention as such the principle of relativity and the postulate of 
the constant speed
of light in vacuum  in special relativity, the postulates of 
quantum mechanics defining the
wavefunction and the observables; further, on somewhat a different 
level, the de Broglie's
periodic phenomenon, the indistinguishability principle, and so on.
Another problem is connected with a lack of consistency in introducing such fundamental
notions as space, time, velocity.
For instance, the concept of a unified world time, or cosmic time 
in cosmology which is
necessary to provide the metrics of the space to be the same within 
the universe, seems
quite different from the locally measured time in special relativity 
which is dependent upon
the choice of a reference system.
Also the role of  Hubble's law, which  is agreed to be of general 
importance in cosmology
precisely for relatively ``small" distances, is not considered  in the course
of introduction of space, time and velocity on the local level.
Of course,
physicists are not happy about such a wide diversity of principles, and there
is a continuous and long lasting striving for the unification of the basic
principles in order to reduce their number, or  at least to straighten out the
existing set of fundamentals in order to find out the ``most fundamental"
principles, if not a single principle.
In doing so,  a quite natural question
arises: to what extent does the actually existing variety of fundamental
principles reflect the basic laws and to what extent has it been influenced by
the historical development of physics?
Say, would Albert Einstein formulate
his fundamentals of special relativity in the same way if the de Broglie's
periodic phenomenon,  the indistinguishability principle and Hubble's 
law would
have already been widely accepted in that time?

Let us remind that Einstein's special relativity theory is based on two concepts, the inertial
system of reference, and the principle of relativity.
It is clear however that, prior to the introduction of the inertial system of reference, one
has to start with introduction of a system of reference in general.
In this connection it is worth to cite the very first statement opening the presentation of the
special relativity theory in the  widely accepted Course of theoretical physics by Landau
and Lifshitz \cite{LL-86}: ``For the description of the {\it processes taking place in nature}, one
must have a {\it system of reference}.
By a  system of reference we understand a system of coordinates serving to indicate the
position of a particle in space, as well as clocks {\it fixed} in this system serving to indicate the
time" (emphasis added, R.F.).
The existence of  inertial frames is then defined for a {\it freely moving body (entity)}
which proceeds with constant arbitrary velocity.
``Suppose that in a certain inertial system we observe clocks which is moving relative to
us. ...
Thus at each moment of time we can introduce a coordinate system rigidly linked to the
moving clocks, which with the clocks constitutes an inertial reference system" \cite{LL-86}, such
clocks showing the proper time in this system.
One can clearly see a contradiction in the above statements: the operation of {\it
rigid linking} necessarily involves {\it interaction}, which is, by definition, excluded for a
{\it free} entity. 
And it is quite unclear ``who" and ``how" will provide
``each free entity" with a clock, a ruler and a radar since it seems dubious to
imagine cumbersome constructions attached to a free entity which can be, say, a
fundamental particle.
At the same time, according to the very idea of
relativity, {\it each} free object (entity) may be chosen as the center of an
attached system of reference.
The last but not least point in fundamentals of
special relativity is connected with the {\it velocity} notion, which enters
the postulate defining inertial reference frames, whilst the velocity notion
itself is defined only in linkage with moving reference frames.

The foregoing discussion casts some doubt upon actual existence of inertial reference
frames {\it in nature} and seems to be the weakest point in commonly agreed fundamentals
of the relativity theory, allowing one to suspect that the theory itself is self-contradictory.
What is more, it is not excluded that the above-mentioned lacunas in formulation of special
relativity are responsible, at least to a certain extent, for existing difficulties with
interpretation of quantum mechanics, since the latter had appeared as a result of merging
of special relativity with quantum postulates.

The present paper presents an attempt to suggest an alternative way of looking at the
fundamentals of relativity.
The following basic idea is proposed.

Each free entity (object) is defining a system of reference {\it by itself}, providing the ideal
{\it internal standards} of time and distance, and also of velocity, without any need of any
other object to be attached to.
Since the only intrinsic characteristic of a free entity is its {\it proper mass}, this
requirement is actually implemented by providing internal standards which are defined
exclusively by the entity's proper mass and the world constants.
Hence, I suggest that the internal inconsistency in fundamentals of relativity may be
overcome by restating the above cited postulate \cite{LL-86} introducing the system of reference.
The suggested formulation is as follows: {\it processes have to take place in nature in
such a way, which provides an actual existence of internal standards of time, distance and
velocity for each free entity defined by its proper mass}.
In this sense, not only is each free entity (single object) {\it defining} a system of
reference, but is {\it itself defined} by the latter.
In slightly different words, the proposed approach states that to postulate the {\it actual
existence} of reference frames means to postulate the {\it actual existence} of internal
time, distance and velocity standards and this, in its turn, means to define the {\it actual
existence} of the entity {\it in reality}.
The foregoing proves that to formulate the fundamentals of relativity means to formulate a
set of basic physical laws (principles, phenomena) which, as taken together, will ensure the
realization of the above postulate.

\section{Internal Time Standards}

We have to start with a {\it proper time} standard in entity's proper reference frame.
Indeed, it is easy to understand that by the very definition of a proper reference frame its
{\it only} intrinsic characteristic is the {\it proper time t} because any spatial coordinates
are zero and therefore velocity cannot be introduced.
From another point, since the {\it only} intrinsic characteristic of each free entity, which is
not related to any other object or system of reference, is its proper mass $m_0$, the
latter could be considered as necessary and sufficient condition of existence of an object
(entity) which can represent a proper system.
Hence, the only possible way of existence of {\it internal standards}, in particular, of time,
is that any proper system of reference characterized by its {\it proper mass $m_0$} has
inevitably to be the ideal clock {\it by itself}.

The latter statement that the entity is {\it by itself} functioning (operating) as a clock does
in reality mean that we have to define both the standard internal time {\it periods $T_0$}
and the standard internal procedure of {\it measuring} the lapse of time $t$.
This can be formulated as the following requirements:

(1) {\it standard proper time periods $T_0$} have to exist for each free entity, and they
have to be fully determined exclusively by the entity's {\it proper mass $m_0$};

(2) {\it counting of periods}: functioning ``as a clock" needs an operation which is nothing
else but uninterrupted, continuous coherent {\it counting} of standard proper time periods
$T_0$;

(3) {\it ideal clocks}: since the ``clocks" have to be {\it ideal}, both the counting
operation and the time standard $T_0$ have to be {\it ideal}, the latter meaning that the
proper mass $m_0$ has to be, at least in principle, defined with {\it absolute} precision,
that  is as good as is defined any other physical constant, for instance, the constant
expressing $T_0$ via $m_0$.

Let us analyze, how the processes have to take place {\it in nature} in order to provide the
fulfillment of the above-enumerated requirements.

\subsection{Standard proper time periods.}

Such periods $T_0$ determined by the entity's proper mass $m_0$ immediately bring
into existence the {\it de Broglie's internal periodic phenomenon}.
It is interesting to remind that de Broglie himself had an idea of some ``great" law of
nature standing behind it \cite{Brogl}: ``One may therefore assume that, {\it as a result of a great
law of nature} every bit of energy of proper mass $m_0$ is intrinsically related to a {\it
periodic phenomenon} of frequency $\nu_{0}$ in such a way that $h \nu_{0} =
m_{0}c^2$, $\nu_{0}$ being, evidently, measured in the system {\it attached} to a bit of
energy.
This hypothesis is the basis of our system: it is valid, like any hypothesis, as much as are
valid the consequences which one can deduce from it" (emphasis added, R.F.).
It is easy to see that {\it de Broglie's periodic phenomenon} introduces for each entity the
time periods $T_{0} = 1/\nu_{0} = (h/c^{2}) m_{0}^{-1}$ which are indeed the desired
{\it time standards} being defined exclusively by the entity's proper mass $m_0$.
Thus, an internal time standard is introduced  by the internal de Broglie's periodic
phenomenon in the most natural and fundamental way, the phenomenon being the only
intrinsic characteristic necessary for the very existence of a free entity.

\subsection{Counting of periods.}

It is necessary to define not only a time standard, but the measurement procedure as well.
It is clear that the operation of a clock needs the ideal periodic process.
Mathematically this operation can be represented by a simplest periodic function of the
type, say, $sin(2\pi t /T_{0}) = sin[(2\pi /T_{0})(\tau + n T_{0})]$, or $e^{2\pi i (\tau + n T_{0}) 
/T_{0}}$, which can
be associated with the counting of periods $T_0$ in the course of ongoing (lasting)
time $t$ with $n$ going on as 1, 2, ... .
The continuous proper time lapse $\tau$ has then to be ``inside" the period, $\tau \le
T_{0}$, which is, by definition, below the measurement limit and, thus, is immeasurable in
concept.

Let us dwell on the meaning of the ``immeasurable" time lapse $\tau$.
First, its important role is to ensure the continuity of lasting time $t= \tau+ nT_{0}$.
Next, the fact of being {\it conceptually} ``immeasurable" means that it is impossible to
define the measurement procedure for this purpose.
With this in mind, one may consider that $\tau$ is {\it conceptually  undefined}, and the
maximal relative accuracy of time measurement, for a given time interval $t$ which
corresponds to a fixed $n$, is equal to $T_{0}/t = 1/(1+n) \simeq 1/n$.
This correlates nicely with the usual idea of the accuracy dependence on the number of
measurements, thus supporting the constructive definition of the internal measurement
procedure.
Speaking about the fundamental inaccuracy of time measurement, it can be immediately
judged that, since $\tau \le T_{0}$, the maximal inaccuracy $\Delta t$ is equal to one
period, $\Delta t = T_{0} = h/(m_{0}c^{2})$, which yields $\Delta t \times (m_{0} c^{2})
= h$, and one arrives, in the most evident and simplest way, at the {\it uncertainty
relation} $\Delta t \times \Delta E = h$.
Here one gains an important insight into the meaning of the rest energy, $\Delta E =
m_{0} c^{2}$, which may be considered as being conceptually  {\it undefined} during the
lapse of time $\tau \le T_{0}$ within which the entity's proper time is {\it undefined}.
Or, equally, the larger the entity's proper mass, the shorter the time lapse within which the
very existence of the entity itself is {\it undefined}, that is, so to speak, within which the
entity has not to be conserved.
In other terms, the phenomenon of fluctuations of the physical vacuum has been brought
into existence.

Let us now try to disclose the essence of the counting procedure.
The important point is that it is {\it time itself} which adds a unit to $n$, passing to $n +
1$.
As a result, the very idea of such counting may be understood only in close linkage with
the interpretation of the concept of time itself.
As proposed earlier in Ref.~\cite{ferber}, the proper time evolution of the entity is actually realized in
such a way that, in the course of time, $n$ runs coherently through all integer numbers.
This allows to suggest that the {\it numbers} themselves are {\it realized} in the course of
counting periods with running time.
As it follows from the above-mentioned, $t/T_{0} \sim t m_{0}$ and, from another point of
view, $t/T_{0} \sim n$, which means that $n \sim t m_{0}$, thus showing that the
``running" of natural numbers is ensured by the ``running" time scaled by means of the
proper mass.
Hence, involving {\it periodicity} and {\it time course} means involving {\it positive
integers, or numbers}, and can be therefore considered as the {\it arithmetic concept of
time} or, equivalently, the {\it physical realization of the set of natural numbers} which is
known as the basic object of arithmetic.

It is worth adding that the proper time ``measurement" does not mean that something is
measured by any external observer, conscious or not, but is rather {\it self-measured}
(which is equal to self-defined).
Since there is no observer for a really and absolutely free entity, we have to attribute self-measuring 
with numbering, and proper time is then a numbered (ordered) sequence of
periods, $t = T_{0} , 2T_{0} , 3T_{0} , ..., nT_{0} , ...,$ which can be associated with the
self-evolution of the entity.

\subsection{Ideal clocks.}

The next arising problem is how to ensure the clock's {\it ideality}.
This requirement, as follows from the above discussion, includes two points.
First, since in order to ensure {\it ideality} of entity's proper reference system time
standards $T_{0}$ one has to ensure the {\it ``ideality"} of the entity's proper mass
$m_{0}$, the latter has to be known, at least in principle, with ideal (that is, absolute)
precision.
As was stressed in Ref.~\cite{ferber}, the only way to make it possible is to formulate the {\it
principle of indistinguishability}.
Indeed, owing to such a principle one may be sure about the actual existence in nature of
{\it ideal} time standards which can be fixed by fixing the $m_{0}$ value: even the slightest
difference in proper mass of, say, two free electrons would allow to distinguish them.
And, what is more, we arrive here at a very important point, namely at the definition of a
notion of a {\it really free} entity (object): such entities have to be indistinguishable as are,
for instance, the fundamental particles.

Secondly, the {\it ideal} internal clocks require the {\it ideal} internal counting procedure.
As mentioned above, the procedure of counting through measuring time is carried out
according to the scheme of natural number progression (by natural numbers I mean the 1,
2, 3, ... progression, excluding zero).
The progression of integers is the only mathematical idealization of the procedure of real
counting and as such possesses the power of an only possible ``absolute" truth, on which
the tremendous power of mathematics in exact sciences is based.
It is considered that the absolute power of the progression of integers is extended also to
physics, making, in principle, the creation of physical theories possible.
The preceding discussion may possibly permit to shed a different light upon such a
surprising power: indeed, the basic object of arithmetic are the integers, as are the free
fundamental entities in physics, and the very existence (self-existence) of such entities {\it
in nature} needs, or can be associated with, the progression of integers.
Let us also note that the famous negative results of Kurt G\"odel are based on 
an assumption that in the course of construction of mathematical formulae for a 
fully formalized physical theory the principles of counting and ordering of 
formulae never change, obeying the laws of natural numbers even for very large 
sets of formulae.  
It is however a long lasting problem of actual realization 
of {\it natural numbers} in nature.  
The notion of natural numbers, which 
consistently defies any definition, was usually connected with counting of 
ideal objects, leading to the question, whether an entity (object) may exist in 
infinitely large numbers of its absolutely identical copies.  
Since one can not 
distinguish between the ``same" and ``another" fundamentally indistuingishable 
objects, it is conceptually impossible to distinguish between the counting of 
{\it periods} of the {\it same} entity and counting of {\it copies} of the 
indistuingishable objects.  
This offers a possibility of the existence 
(generating) of infinite number of absolutely identical ``copies" since 
periodicity may be understood as the ordered sequence of reproduction of the 
entity (or indistuingishable entities) after equal time periods $T_{0}$.  
And, finally, one more note.  
As became known in recent years, the power of {\it 
quantum computing} is going to be, at least for a curtain class of problems, 
extremely and somewhat unexpected high owing to parallel employment of a large 
number of absolutely identical units.

Summarizing the Section, it is possible to conclude that, had the {\it de Broglie's internal
periodic phenomenon} and the {\it principle of indistinguishability} not existed, they
would have had to be introduced as the fundamentals of the special relativity theory in
order to avoid the internal self-contradiction in its foundations.

\section{Distance and Velocity Standards}

If we are starting from proper time, the most difficult point is how space should be
introduced.
In the frame of the present paper we have, at first, to admit that {\it if} the reference frame
is associated with a free entity, defined as {\it absolutely ideal clocks} (on the proper time
level the entity is only ``clocks", with no other identification) the same entity has inevitably
to be {\it by itself} also {\it the absolutely ideal ``ruler"} to define (self-measure) the distance.

Further, the definition of inertial reference frames involves, besides time and distance, also
the {\it velocity} concept.
In consistency with the approach to fundamentals of special relativity developed here, it is
necessary to postulate that the entity has {\it by itself} to be the {\it absolutely ideal
``radar"} to define (self-measure) the velocity and to make the actual existence of  inertial
reference frames {\it in nature} possible.

Both above requirements, similar to time self-measuring, also demand the existence of
distance and velocity standards, which are defined exclusively by proper mass and world
constants, with the same ideal counting procedure as described above.
Let us search for the existing fundamentals in physics allowing  to fulfill these
requirements.

\subsection{Distance standards.}
A search for an internal {\it distance standard} $\Lambda_{0}$ which would be in
consistency with our approach, is based on the necessity to be fully defined by the internal
time standard $T_{0}  = (h/c^{2})(m_{0}^{-1})$, which means that $\Lambda_{0}$ 
and $T_{0}$ have to be connected via an essentially constant scaling factor.  
These requirements immediately bring the following simplest outcome: $\Lambda_{0}=
cT_{0}$, $\Lambda_{0}$ being known as the Compton wavelength.
Note, that the only requirement for the scaling factor $c$ is to be a world constant (with
velocity dimension), without a necessity to imply any more specific physical  meaning.
Thus, the distance $x$ is constructively introduced by the internal {\it distance standard}
$\Lambda_{0} = cT_{0} = (h/c)(m_{0}^{-1})$ which is again provided by the free entity's
proper mass, avoiding any necessity to imagine any extra attached device, or to attach a
ruler, for the purpose of  measuring (self-measuring) the distance $x$.
And let us stress once again that, in order to ensure standard's ideality, the scaling factor
$c$ has to possess a constant value independently of any internal or external factor, which
may be, from another point, considered as the definition of $c$.

\subsection{Velocity standards.}
Now the question is: how can the internal {\it velocity standard} be introduced?
In search of such a standard I propose to use Hubble's law according to which space and
velocity could not be independent and are intrinsically linked to each other via a scaling
factor.
Let us remind that Hubble's law reflects the expansion of the universe, $x = H_{0}^{-1}
v$, $H_{0}^{-1}$ being Hubble's time, $x$ being the distance between two entities, and
$v$ the receding velocity.
Hubble's law holds locally for any cosmological model, see Ref.~\cite{LL-86}, and may be considered
as a scaling law due to which velocity $v$ is connected with distance $x$ via Hubble's
constant $H_{0}$ as $v = H_{0} x$.
Since the distance standard $\Lambda_{0}$ is defined as $cT_{0}$, this opens the simplest
possibility of defining the {\it internal velocity standard}: $V_{0} = cH_{0} T_{0} =
(H_{0} h/c)(m_{0}^{-1})$.
Similarly, it has to be assumed that the scaling factor $H_{0}$ is a world constant.
Let us stress once more that such a definition is possible only owing to the fact that,
dealing with {\it internal standards}, we are dealing with local ``measurements" when the
Hubble's law is for sure valid.
And it is important that the velocity standard is defined in a {\it non-differential} way by
means of the scaling law.

As distinct from time and distance standards, the introduction of the velocity standard
needs more comments.
The adoption of Hubble's law is based on a formal analogy between connecting time
with distance, $x = ct$, in special relativity and connecting velocity with distance in the
course of the expansion of universe, $x = H_{0}^{-1} v$, as well as on the dimension of
the  respective constants (scaling factors): $[c] = distance/time$ and  $[H_{0}^{-1}] =
distance/velocity$.
Using this analogy, Moshe Carmeli \cite{carm} had recently developed a new formulation of
special relativity, a {\it cosmological special relativity}, which is built upon two
assumptions which had been raised to postulates, namely:
(a) the constancy of expansion of universe expressed by Hubble's law, according to which
$H_{0}$ is constant in all (cosmic) times;
(b) the principle of cosmological relativity which, analogous to the principle of special
relativity, postulates that all physical laws are the same at all cosmic times (as at all inertial
reference frames in special relativity).
The constant $H_{0}$ is considered to be a world constant just as in special relativity the
constant $c$ is considered not to change its value because of gravity.
As a consequence, a transformation has been derived in \cite{carm} relating {\it distances and
velocities} measured at different (cosmic) times, this transformation being an analogue to
the Lorentz transformation relating {\it distance and time}.
Then, a null-vector $x^{2} - (H_{0}^{-1})^{2} v^{2} = 0$ has to be conserved, for any
distances $x$ and $x'$ in any cosmic time $t$ and $t'$, in a zero curvature universe, just as
$x^{2} - c^{2} t^{2} = 0$ has to be conserved in special relativity for Minkowskian space-time.
This leads to the following transformations:
\begin{equation}
x = (x - tv)/ (1 - t^{2}/H_{0}^{-2})^{1/2}, \, \,     x = (v - xt)/(1 - t^{2}/H_{0}^{-2})^{1/2}
,	\nonumber
\end{equation}
$t$ being the cosmic time which is defined backward with respect to the present time.
It is clear that the homogeneity and isotropy of space mean that we can choose a so-called
cosmic time, or such world time \cite{LL-86} which will provide the same metric of space for all points
and directions.
Thus, the roles of time and velocity have been exchanged in the above formulae as
compared to the Lorentz transformations in special relativity, the fundamental variable $t$
being replaced by $v$ and the ratio $v/c$ by $t/H_{0}^{-1}$.
In a sense, similar to the fact that velocity is restricted by $c$ in special relativity, the cosmic
time is restricted by $H_{0}^{-1}$.
The physical consequences of Carmeli transformations \cite{carm} is, among others, that if one
measures different distances and velocities at different (cosmic) times in past, one obtains
{\it length contraction} and {\it velocity contraction} with the factor $(1 - t^{2}/H_{0}^{-
2})^{1/2}$ like {\it length contraction} and {\it time dilation} with the factor $(1 -
v^{2}/c^{2})^{1/2}$ in special relativity.

Concluding this Section, it is worth mentioning that to provide distance and velocity
standards defined exclusively by the free entity's proper mass we have exploited, besides
de Broglie's periodic phenomenon, the Hubble's law resulting in the same symmetry
between expressions connecting time with distance, $x = ct$, and distance with velocity,
$x = H_{0}^{-1} v$, which had permitted Carmeli to formulate the cosmological special
relativity.
Owing to such a symmetry, leading to the similar form of transformation between reference
systems, it becomes possible to introduce a non-differential definition of the velocity
concept.

\section{Consistent Relativity and Basic Laws}

This Section is aimed to summarize the preceding subject and to sketch briefly the contents
of further discussion, concerning, in particular, the application of the developed approach
to interpretation of quantum physics \cite{ferb2}.

\subsection{Consistent relativity.}
As demonstrated, the consistent fundamentals of special relativity are the basic laws of
nature necessary to ensure actual existence of proper systems of reference by introducing
internal time, distance and velocity standards defined exclusively by the entity's proper
mass.
Summarizing the definitions of all three internal standards, we obtain:

(1) {\it time standards},          $T_{0} = (h/c^{2})(m_{0}^{-1}) \equiv  k_{T}/m_{0}$ ,

(2) {\it distance standards},    $\Lambda_{0} = c T_{0} = (h/c)(m_{0}^{-1}) \equiv
k_{\Lambda}/m_{0}$,

(3) {\it velocity standards},    $V_{0} =  cH_{0} T_{0} = (H_{0} h/c)(m_{0}^{-1})\equiv
k_{V}/m_{0}$.

It is easy to notice that, if one starts with the time standard, which is defined by reversed
proper mass via the factor $k_{T} \equiv h/c^{2}$, then the factors defining two other
standards can be obtained by subsequently multiplying $k_{T}$ with $c$ yielding
$k_{\Lambda} \equiv h/c$ (note that defining $c$ means that both world constants $c$
and $h \equiv k_{T}  c^{2}$ appear separately) and then with $H_{0}$ yielding $k_{V}
\equiv H_{0}h/c$.
Let us stress that the hierarchy between the standards according to which everything can
be obtained from the proper time standards fits well Einstein's general relativity theory in
which the proper time is used to define the invariant separation between two events,
allowing to derive every other physical concept (spatial intervals, etc.) \cite{LL-86}.
It is also important to mention  that time, distance and velocity standards (1) - (3) appear in
a symmetric way whilst time and space only are the fundamental variables entering the
Lorentz transformation in the special relativity theory which includes velocity as a
parameter.

Hence, in order to provide the internal {\it time, distance} and {\it velocity standards}, or,
equally, to ensure the {\it consistent} fundamentals of special relativity, the processes
inevitably {\it have to} ``take place in nature" according to de Broglie's periodic
phenomenon and Hubble's law.
To provide the {\it ideal} standards, the {\it principle of indistinguishability} should exist,
and $c$ and $H_{0}$ {\it have to} be world constants.
These laws are thus forming the {\it consistent basis} of the relativity theory by defining a
proper reference frame for an entity with a definite proper mass, without any linking of
external chronometers, rulers or radars.
Further, the actual existence of each free object (entity) means actual realization of ideal
``clocks" operation as actual realization of a regular progression of integers which is
``numbering", or counting de Broglie's periods $T_{0}$ in the course of time, allowing to
link the (local) time notion with physical realization of  natural numbers \cite{ferber}.
This opens the way out of the vicious circle in Einstein's relativity theory, where a clock is
needed to test the dynamical laws on which its (the clock's) functioning itself is based.

\subsection{Consistent relativity and Hubble's law.}
Including Hubble's law amongst the fundamentals of relativity fits the idea that the
processes in nature at the {\it global} (cosmological) scale  have to proceed in such a way
as to provide the realization of a system of reference at the {\it local} scale.
This is a step towards meeting the objective of consistency between physics in the small
and large scales.
Hubble's law is necessary for passing from the (proper) time to space and velocity: in a
sense, ``creation" of space and velocity is realized at the {\it local} scale in the course of
expansion of the universe at the {\it global} scale.
It is important to stress that, as follows from comparing Carmeli \cite{carm} and Lorentz
transformations containing time and velocity as respective parameters, this is taking place
{\it backward} in time with respect to the present (cosmic) time moment $t = 0$, which is
a ``preferred" time moment just as $v = 0$ can be regarded as ``preferred" velocity in
proper system.
This brings to one's mind the idea that, for an isotropic space--time model of the universe
in relativistic cosmology \cite{LL-86}, it is most convenient to choose a reference system which is
``co-moving" with the ``matter", such reference system being just the matter ``filling" the
space, the {\it velocity} of the ``matter" being zero everywhere by definition.
Thus, as distinct from proper time standards, velocity (and distance) standards are defined
for the space which has already been ``created" by expanding in the past in the course of
expansion.
Space and velocity have already {\it existed}, or have been {\it created}, at all cosmic
times in the past with respect to the present time.

\subsection{Consistent relativity and quantum physics.}
This problem deserves, of course, a separate discussion \cite{ferb2} and will be touched upon here
only very briefly.
It is commonly known that the merging of special relativity theory with quantum
mechanics has led to most advanced understanding of atomic and subatomic physics, as
well as of electromagnetism in general.
I would however suggest a different statement: {\it consistent relativity based upon
internal standards inevitably leads to quantum mechanics}.
In the course of the discussion in the previous sections, a quite natural question may arise:
how to imagine  (to interpret) the ``internal" system of reference which is intrinsically
associated with an entity? 
I would say that this is exactly as difficult as is to imagine (to
interpret) the {\it wave function}.
And if so, it is probably a good way to consider the wave function as an internal system of
reference which is ``probing", or ``scaling" the space--time.
Let us look for the arguments in favour of such an approach.
For a free entity (particle), both the wave function in quantum mechanics and the system of
reference with internal time standards in special relativity stem from one source, namely
from the de Broglie's periodic phenomenon.
Indeed, let us remind \cite{Brogl,ferber} that since the dimensionless phase of de Broglie's periods has
to be conserved under Lorentz transformation of proper time $t' = (t - xv/c^{2})/(1 -
v^{2}/c^{2})^{1/2}$, applying this transformation to the phase of de Broglie's periods,
immediately yields the expression which equally represents the de Broglie's waves of
matter and the wave function of a free particle $\Psi (\omega_{0}  t') = \Psi [\omega_{0}(t
- xv/c^{2})/(1 - v^{2}/c^{2})^{1/2}] \equiv  \Psi (\omega t - kx)$.
And from the very feature of wave motion, Heisenberg's uncertainty principle in general
and, for instance, the discreteness of energy levels for a bonded system (atom, molecule)
follow in a direct way.
Now, de Broglie's waves can be considered as the solution of the Klein--Gordon equation
which, in non-relativistic limit, agrees with the Schr\"odinger equation.
Thus, the entire quantum mechanics is based on the de Broglie's periodic phenomenon,
and it can as appropriately be concluded that the necessity to introduce the {\it wave
function} follows from the necessity to provide the {\it internal time standards} or, more
generally, a reference frame in consistent relativity theory.

Further, it is worth noting that, as is agreed in quantum mechanics, the {\it wave function}
is introduced as a function by which an object (in particular, a free particle) is completely
{\it defined}, and this exactly corresponds to the conclusion made in Section I that each
free entity is {\it defined} as a single object by an {\it internal system of reference}.
In this sense, both the {\it internal system of reference} and the {\it wave function}
(represented in space--time coordinates) can be considered as an internal feature
intrinsically linked with each entity (object), defining the latter and serving to describe its
actual existence and behavior in space and time.
And what else does the idea of avoiding any external clocks and rulers by introducing some
internal proper {\it system of reference} mean, if not introducing an internal proper {\it
wave function}, the latter being a typical relativistic notion emerging as a direct
consequence of the Lorentz transformation of the de Broglie's oscillations?

If the wave function of a free particle can be associated with internal inertial frame, one has
to consider its ``deformation", when any kind of interaction (external field, etc.) is included
and the frame becomes non-inertial.
This brings to one's mind the validity of curved space--time in general relativity.
It is probably worth noting that some results obtained on the basis of general relativity may
seem controversial due to implicit assumptions of fixing the reference frame.
To make the coordinate representation of the results of general relativity consistent, the
system of reference has to be ``hitched to" the space--time events in a flexible way.
And the idea of identifying the distortion of the wave function caused by interaction, with
the distortion of the internal reference frame seems to follow quite naturally from the
context of the present discussion.
Again, this agrees with the main idea of general relativity, namely that everything can (or
has to) be derived from the proper time interval which is measured (``scaled") by the
proper ``ideal clocks".
What is more, it is by no means accidental that the time standards are defined by nothing
else than the mass, the notion which is the basic quantity in general relativity.


\begin{references}

\bibitem{LL-86} L. D. Landau and E. M. Lifshitz, {\it The classical theory of fields}  (Butterworth-
Heinemann, Oxford, 1996).
\bibitem{Brogl} Louis de Broglie, {\it Recherches sur la theorie des quanta}, Ann.\ Phys. {\bf 3}, 22-
128 (1925).
\bibitem{ferber} R. Ferber, {\it A Missing Link: What is Behind de Broglie's ``Periodic 
Phenomenon"?},
Foundations of Physics Letters {\bf 9}, 575-586 (1996).
\bibitem{carm} M. Carmeli, {\it Cosmological Special Relativity: A Special Relativity for Cosmology},
Foundations of Physics {\bf 25}, 1029-1040 (1995); {\it Cosmological Special Relativity}, ibid. {\bf 
26}, 413-416 (1996).
\bibitem{ferb2} R. Ferber, {\it Concistent Relativity and Interpretation of Quantum Mechanics}, to be
published.

\end{references}
\end{document}